\documentclass[prl,twocolumn,amsmath,longbibliography]{revtex4-1}
\usepackage{graphicx,epstopdf}
\usepackage{bm}
\newcommand{\bracket}[1]{\langle#1\rangle}
\newcommand{\ket}[1]{|#1\rangle}

\DeclareMathOperator{\RE}{Re}

\DeclareMathOperator{\TR}{Tr}

\usepackage[colorlinks=true, letterpaper=true, pdfstartview=FitV,
  linkcolor=blue, citecolor=blue, urlcolor=blue]{hyperref}

\begin{document}

\title{Tunable Layer Circular Photogalvanic Effect in Twisted Bilayers}

\author{Yang Gao}

\affiliation{Department of Physics, Carnegie Mellon University,
  Pittsburgh, PA 15213, USA}

\author{Yinhan Zhang}
\affiliation{Department of Physics, Carnegie Mellon University,
  Pittsburgh, PA 15213, USA}

\author{Di Xiao}

\affiliation{Department of Physics, Carnegie Mellon University,
  Pittsburgh, PA 15213, USA}

\date{\today}

\begin{abstract}
We develop a general theory of the layer circular photogalvanic effect (LCPGE) in quasi two-dimensional chiral bilayers, which refers to the appearance of a polarization-dependent, out-of-plane dipole moment induced by circularly polarized light.  We elucidate the geometric origin of the LCPGE as two types of interlayer coordinate shift weighted by the quantum metric tensor and the Berry curvature, respectively.  As a concrete example, we calculate the LCPGE in twisted bilayer graphene, and find that it exhibits a resonance peak whose frequency can be tuned from visible to infrared as the twisting angle varies.  The LCPGE thus provides a promising route towards frequency-sensitive, circularly-polarized light detection, particularly in the infrared range.
\end{abstract}

\maketitle

Recent years have seen a surge of interest in twisted van der Waals heterostructures consisting of atomically thin crystal layers.  From a structural point of view, twisted layers are interesting because not only are they chiral, but their chirality can be readily controlled by varying the twisting angle~\cite{Kim2016}.  For example, bilayers with opposite twisting angles are mirror images of each other, therefore they possess opposite chirality.  This structural flexibility makes twisted van der Waals heterostructures a versatile platform to investigate chirality-dependent phenomena.  One such example is the surprisingly strong circular dichroism reported in twisted bilayer graphene at large twisting angles~\cite{Kim2016,Morell2017,Stauber2018,Stauber2018a,Addison2019}.

\begin{figure}
  \includegraphics[width=\columnwidth]{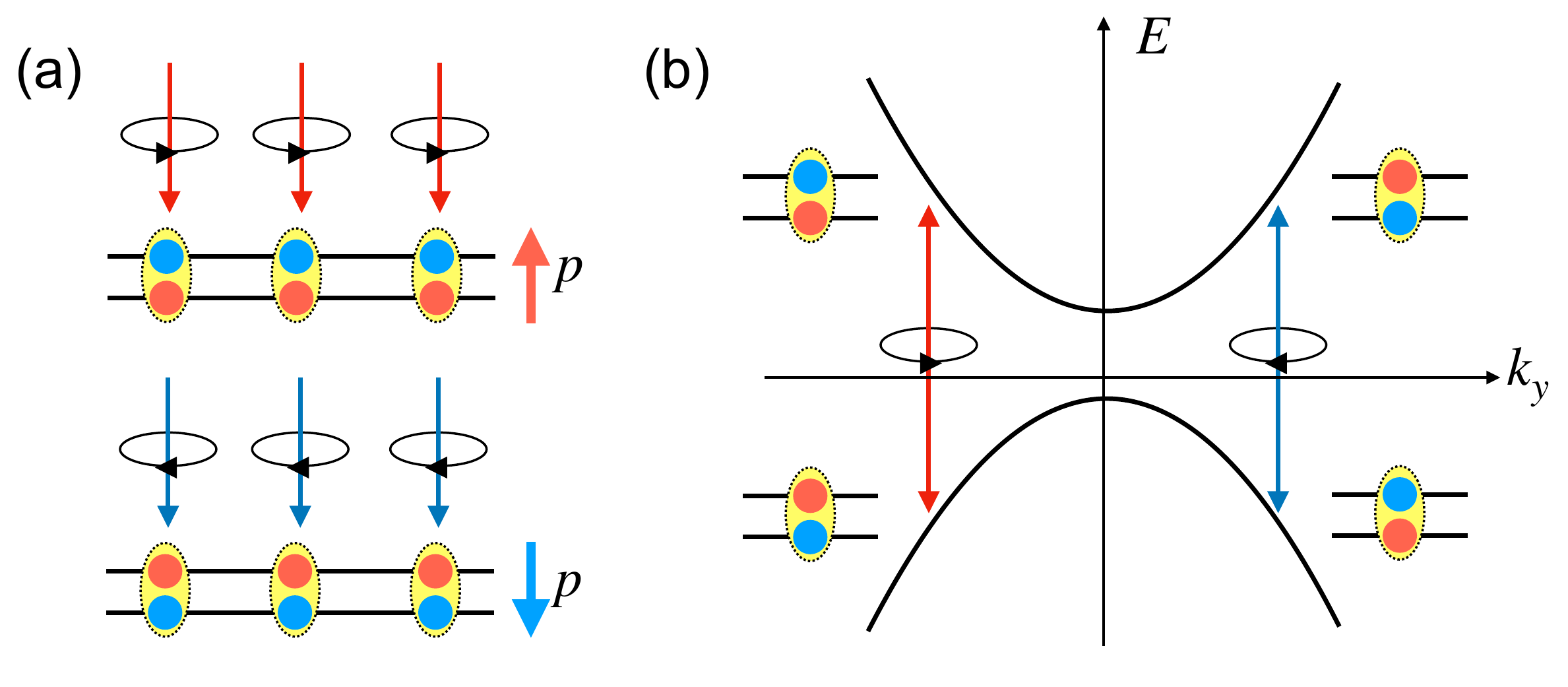}
  \caption{(a) Schematic illustration of the LCPGE. Circularly polarized light induces a static out-of-plane dipole, whose direction flips when the circular polarization of light is reversed. Red and blue disks stand for negative and positive charges, respectively. (b) Origin of the LCPGE.  The incident light excites an electron from the valence band to the conduction band, while simultaneously causes a change in the dipole moment.  Such a transition process is dependent on the light chirality through the geometric factor $\beta_G$ in Eq.~\eqref{eq_geo}.}\label{fig_fig0}
\end{figure}

In this Letter, we explore the consequence of structural chirality of twisted van der Waals bilayers in nonlinear optical effects.  We show that a static, out-of-plane dipole moment can be induced by circularly polarized light at normal incidence [Fig.~\ref{fig_fig0}(a)], which we refer to as the layer circular photogalvanic effect (LCPGE).  We first derive a general expression of the LCPGE coefficient, valid for any quasi two-dimensional chiral bilayers.  The LCPGE has an elegant geometric interpretation: it consists of two types of interlayer coordinate shift, weighted by the quantum metric tensor and the Berry curvature, respectively.  In this regard, the LCPGE is distinctively different from the bulk CPGE~\cite{Moore2010,Deyo2009,Sodemann2015,Juan2017}, and resembles more of the shift current~\cite{Kraut1979,Baltz1981,Sipe2000,Sturman1992,Morimoto2016,Morimoto2016,Yang2017}.

\begin{figure}[t]
\includegraphics[width=\columnwidth]{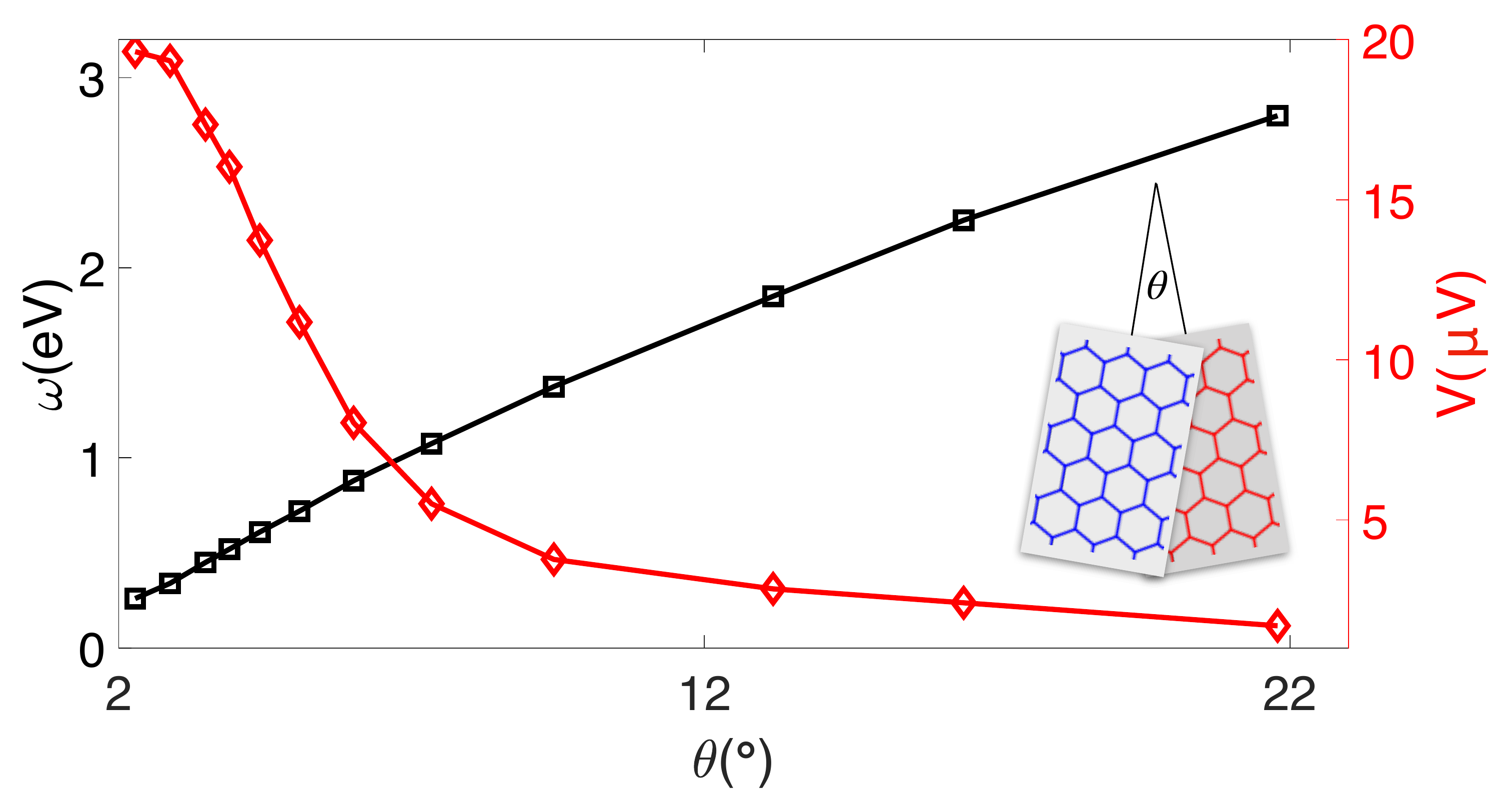}
\caption{The peak position~(black) and the corresponding potential difference~(red) generated by the LCPGE at different twisting angles. We have assumed a laser power 1 mW/$\mu$m$^2$.}
  \label{fig_fig5}
\end{figure}

We then demonstrate the tunability of the LCPGE in twisted bilayer graphene.  We find that the LCPGE signal exhibits a resonance peak determined by three factors: the enhanced density of states, the quantum metric tensor, and the finite interlayer coordinate shift.  The peak frequency can be tuned from visible to infrared with decreasing twisting angle and, at the same time, its magnitude increases sharply (Fig.~\ref{fig_fig5}).  For example, at about 2$^\circ$ twisting angle, the induced voltage difference between the two layers is found to be 20 $\mu$V by a circularly polarized light at 250 meV with a power of 1 mW/$\mu$m$^2$.  These properties make the LCPGE in twisted bilayer graphene a promising candidate towards frequency-sensitive, circularly-polarized light detection in the infrared range.

{\it General Theory.}---Let us consider a generic quasi two-dimensional chiral bilayer, i.e., its structure lacks any mirror plane and inversion center.  The out-of-plane dipole moment is represented by the operator $\hat p = -e\sigma_z$, where the Pauli matrix $\sigma_z$ operates in the layer index subspace and we have set the distance between the two layers to be unity.  We further assume that the system has an in-plane $C_{2x}$ axis, as is the case of twisted bilayer graphene.  The $C_{2x}$ symmetry forbids the existence of the out-of-plane dipole in equilibrium.

Under normal incidence, the dipole $\bracket{\hat p}$ has the following static component
\begin{align}
\bracket{\hat p} =\chi_{ij}(\omega,-\omega) E_i(\omega) E_j(-\omega) \;.
\end{align}
We can decompose the response coefficient $\chi_{ij}$ into a symmetric part $\chi_{ij}^\text{s}$ and an antisymmetric part $\chi_{ij}^\text{as}$.  Since $\hat{p}$ is odd under $C_{2x}$, $\chi_{ij}^\text{s}$ can only have off-diagonal components.  In addition, if there is a more-than-two-fold rotational axis in the $z$-direction, $\chi_{ij}^\text{s}$ has to vanish.  In contrast, $\chi_{ij}^\text{as}$ transforms as a pseudoscalar, i.e., it is invariant under rotation, but flips sign under mirror reflection and space inversion.  Therefore $\chi_{ij}^\text{as}$ is allowed as long as the crystal structure is chiral.

The antisymmetric tensor $\chi_{ij}^\text{as}$ directly couples to the handedness of light.  This can be seen by recasting the polarization $\bracket{\hat p}$ due to $\chi_{ij}^\text{as}$ into the following form
\begin{align}
\langle \hat{p}\rangle=\beta [iE_x(\omega)E_y(-\omega)-iE_y(\omega)E_x(-\omega)]\,,
\end{align}
where $\beta=-i\chi_{xy}^\text{as}$. The expression inside the square bracket is proportional to the fourth Stokes parameter which reflects the circular polarization of light.

The LCPGE coefficient $\beta$ can be obtained as follows.  In the presence of an incident light, the Hamiltonian reads $\hat{H} = \hat{H}_0 + e\hat{\bm v}\cdot \hat{\bm A}$ with $\hat{H}_0$ the unperturbed Hamiltonian and $\bm A$ the vector potential.  We then solve the change to the density matrix up to second order and use it to calculate $\bracket{\hat p}$.  The details are left to the Supplemental Material~\cite{suppl}.  Let $\beta = -e^2\beta_0/\omega^2$; we find that $\beta_0$ is given by a summation over three band indices,
\begin{equation} \label{eq_res1}
\begin{split}
\beta_0 = -i\sum_{\ell,m,n}\int \frac{d\bm k}{(2\pi)^2} & \frac{(v_y)_{m\ell}(v_x)_{\ell n}-(x\leftrightarrow y)}{\omega_{nm}+i\eta_1} \\
&\times (G_{\ell n}+G_{m\ell})p_{nm} \;,
\end{split}
\end{equation}
where $(v_\alpha)_{m\ell}$ and $p_{nm}$ are the velocity and dipole matrix element in the band basis, respectively, $\omega_{nm} = \varepsilon_n - \varepsilon_m$, and $G_{\ell n} = (f_\ell-f_n)/(\omega_{\ell n}-\omega-i\eta_2)-(f_\ell -f_n)/(\omega_{\ell n}+\omega-i\eta_2)$ with $f_n$ being the Fermi-Dirac distribution.  Two phenomenological parameters $\eta_1$ and $\eta_2$ have been introduced to take into account of various relaxation processes.

{\it Geometric origin}.---We now reveal the geometric origin of the LCPGE.  We first show that the intraband contributions ($\varepsilon_n=\varepsilon_m$) in Eq.~\eqref{eq_res1} vanishes.  For nondegenerate bands, $\varepsilon_n=\varepsilon_m$ implies that $n = m$.  In this case, $(v_y)_{m\ell}(v_x)_{\ell m}-(x\leftrightarrow y)$ is proportional to the band-resolved Berry curvature $(\Omega_z)_{m\ell}$ [see Eq.~\eqref{brbc} below], which is odd under time-reversal.  Since both the band energy $\varepsilon_n$ and the dipole moment $p_{mm}$ are even under time-reversal, the integral in Eq.~\eqref{eq_res1} vanishes for intraband contributions.  One can prove that the same conclusion also holds for the degenerate case.

Next we consider the interband contributions ($\varepsilon_n \neq \varepsilon_m$).  If  $\eta_1 \ll |\omega_{nm}|$, we can approximate $\omega_{nm} + i\eta_1 \approx \omega_{nm}$.   It is convenient to introduce an auxiliary Hamiltonian $\hat H_0(\lambda) = \hat H_0 + \lambda \hat{p}$ with $\lambda$ being the layer potential difference.  We can then write the interband dipole moment as $p_{nm} = \bracket{u_n|\hat p|u_m} = \omega_{nm} \bracket{\partial_\lambda u_n|u_m}$, where $e^{i\bm k \cdot \bm r} \ket{u_n(\lambda)}$ is the $\lambda$-dependent Bloch function of $\hat H_0(\lambda)$.  Insert this expression into Eq.~\eqref{eq_res1} and let $\lambda \to 0$ in the end.  We find that, in the clean limit ($\eta_2 \to 0$), $\beta_0$ can be written as a summation over only two band indices~\cite{suppl},
\begin{equation}\label{eq_final}
\beta_0=\lim_{\lambda\to 0} \sum_{\ell,n} \sum_{\xi=\pm 1}\int \frac{d\bm k}{(2\pi)^2} \omega_{\ell n}^2 (f_n-f_\ell)\xi\delta(\omega_{n\ell}+\xi \omega) \beta_\text{G} \;,
\end{equation}
where $\beta_G$ is the geometric factor given by
\begin{equation}\label{eq_geo}
\beta_G= (\TR g_{ij})_{n\ell}(R_\text{as})_{n\ell}-(\Omega_z)_{n\ell}(R_\text{s})_{n\ell}\,.
\end{equation}

We now explain the various terms in $\beta_G$.  The quantities $(\Omega_z)_{n\ell}$ and $(g_{ij})_{n\ell}$, defined by
\begin{align}
(\Omega_z)_{n\ell} &= -2{\rm Im}[\bracket{u_n|i\partial_{k_x}u_\ell}\bracket{u_\ell|i\partial_{k_y}u_n}] \;, \label{brbc} \\
(g_{ij})_{n\ell} &= \RE [\bracket{u_n|i\partial_{k_i}u_\ell}\bracket{u_\ell|i\partial_{k_j}u_n}] \;,
\end{align}
have the meaning of the band-resolved Berry curvature and quantum metric tensor, respectively.  They have the property that summing over one band index will recover the full Berry curvature and quantum metric in the other band, i.e., $\sum_{\ell\neq n}(\Omega_z)_{n\ell}=(\Omega_z)_n$ and $\sum_{\ell\neq n}(g_{ij})_{n\ell}=(g_{ij})_n$~\cite{Provost1980,Berry1984}.

The quantities  $(R_\text{as})_{n\ell}$ and $(R_\text{s})_{n\ell}$ are given by $(R_\text{as,s})_{n\ell}=\frac{1}{2}[(R_+)_{n\ell} \mp (R_-)_{n\ell}]$ with
\begin{align}
(R_\pm)_{n\ell}=\partial_\lambda(\phi_\pm)_{n\ell}-(a_\lambda)_n+(a_\lambda)_\ell\,,
\end{align}
where $(\phi_{\pm})_{n\ell}={\rm arg}(v_\pm)_{n\ell}$ is the phase of the velocity matrix element $v_{\pm}=v_x\pm iv_y$, and $(a_\lambda)_n = \bracket{u_n|i\partial_\lambda|u_n}$ is the Berry connection.  Note that due to the appearance of $a_\lambda$, both $R_{\pm}$ and $R_\text{as,s}$ are gauge-independent.

One immediately recognizes that the expression for $R_\pm$ shares a striking similarity with the coordinate shift in the shift current expression, with the latter given by $\bm R_{n\ell} = \bm \partial_{\bm k}\phi_{n\ell} - (\bm a_{\bm k})_n + (\bm a_{\bm k})_\ell$~\cite{Kraut1979,Baltz1981,Sipe2000,Sturman1992,Morimoto2016}, where $\phi_{n\ell}={\rm arg} (v_x)_{n\ell}$ and $(\bm a_{\bm k})_n=\langle u_n|i\bm \partial_{\bm k}|u_n\rangle$.  Since $\lambda$ is conjugate to the dipole moment $p$, we can interpret $R_\pm$ as the interlayer coordinate shift, which has the desired property that it flips sign under mirror reflection $\mathcal M_z$ with respect to the $xy$-plane.  Thus the LCPGE can be interpreted in terms of geometric quantities defined in the $(\bm k, \lambda)$ parameter space: it consists of two types of interlayer coordinate shifts, weighted by the band resolved quantum metric and the Berry curvature, respectively.

\begin{table}
\caption{\label{tab:transform}Transformation properties of quantities in Eq.~\eqref{eq_geo}.}
\begin{ruledtabular}
\begin{tabular}{l|llll}
 & $\TR g_{ij}$ & $\Omega_z$ & $R_\text{as}$ & $R_\text{s}$ \\
 \hline
$\mathcal M_{x,y}$ & $+$ & $-$ & $-$ & $+$\\
$\mathcal M_z$ & $+$ & $+$ & $-$ & $-$\\
$\mathcal I$ & $+$ & $+$ & $-$ & $-$
\end{tabular}
\end{ruledtabular}
\end{table}

The two-band formula in Eq.~\eqref{eq_final} and Eq.~\eqref{eq_geo} also provides a simple picture of the LCPGE as shown in Fig.~\ref{fig_fig0}(b).  Let us write the integrand of Eq.~\eqref{eq_final} as
\begin{align}
\omega_{\ell n}^2\beta_\text{G}=(W_+)_{n\ell}(p_+)_{n\ell} - (W_-)_{n\ell}(p_-)_{n\ell}\,,
\end{align}
where $(W_\pm)_{n\ell}=\omega_{\ell n}[({\rm Tr}g_{ij})_{n\ell}\mp (\Omega_z)_{n\ell}]$ is nothing but the oscillator strength of the left and right circularly polarized light for the transition from $n$-th to $\ell$-th band~\cite{Souza2008}, and $(p_\pm)_{n\ell} = \omega_{\ell n} (R_\pm)_{n\ell}$ describes the change in the dipole moment that occurs as an electron absorbs a chiral photon.  Therefore the LCPGE directly measures the difference in the induced dipole moment when the electrons are excited by left and right circularly polarized light.

The geometric factor $\beta_\text{G}$ is fully compatible with the point-group symmetry requirement of the LCPGE.  The Berry curvature transforms as a pseudovector, and the trace of the quantum metric tensor $\TR g_{ij}$ transforms as a scalar. The phase factor $\phi_+$ and $\phi_-$ transform in the following way: $\phi_+\to \pi+\phi_-$ and $\phi_-\to \pi+\phi_+$ under $\mathcal M_x$, $\phi_+\to \phi_-$ and $\phi_-\to \phi_+$ under $\mathcal M_y$, and $\phi_+\to \phi_++\pi$ and $\phi_-\to \phi_-+\pi$ under inversion.  Finally, $\partial_\lambda \to -\partial_\lambda$ under $\mathcal M_z$ and inversion.  The transformation properties of the geometric quantities in Eq.~\eqref{eq_geo} are summarized in Table~\ref{tab:transform}.  We can see that $\beta_\text{G}$ is odd under chirality reversal operations such as inversion and mirror operations.

Before closing this section, we wish to remark that even though our LCPGE shares the same symmetry requirement as the out-of-plane component of the bulk CPGE, their geometric origins are completely different.  The bulk CPGE is determined by the Berry curvature dipole~\cite{Moore2010,Deyo2009,Sodemann2015,Juan2017}, while our LCPGE resembles more of the shift current~\cite{Kraut1979,Baltz1981,Sipe2000,Sturman1992,Morimoto2016,Yang2017}, and depends on both the Berry curvature and the quantum metric tensor.

\begin{figure}[t]
\includegraphics[width=\columnwidth]{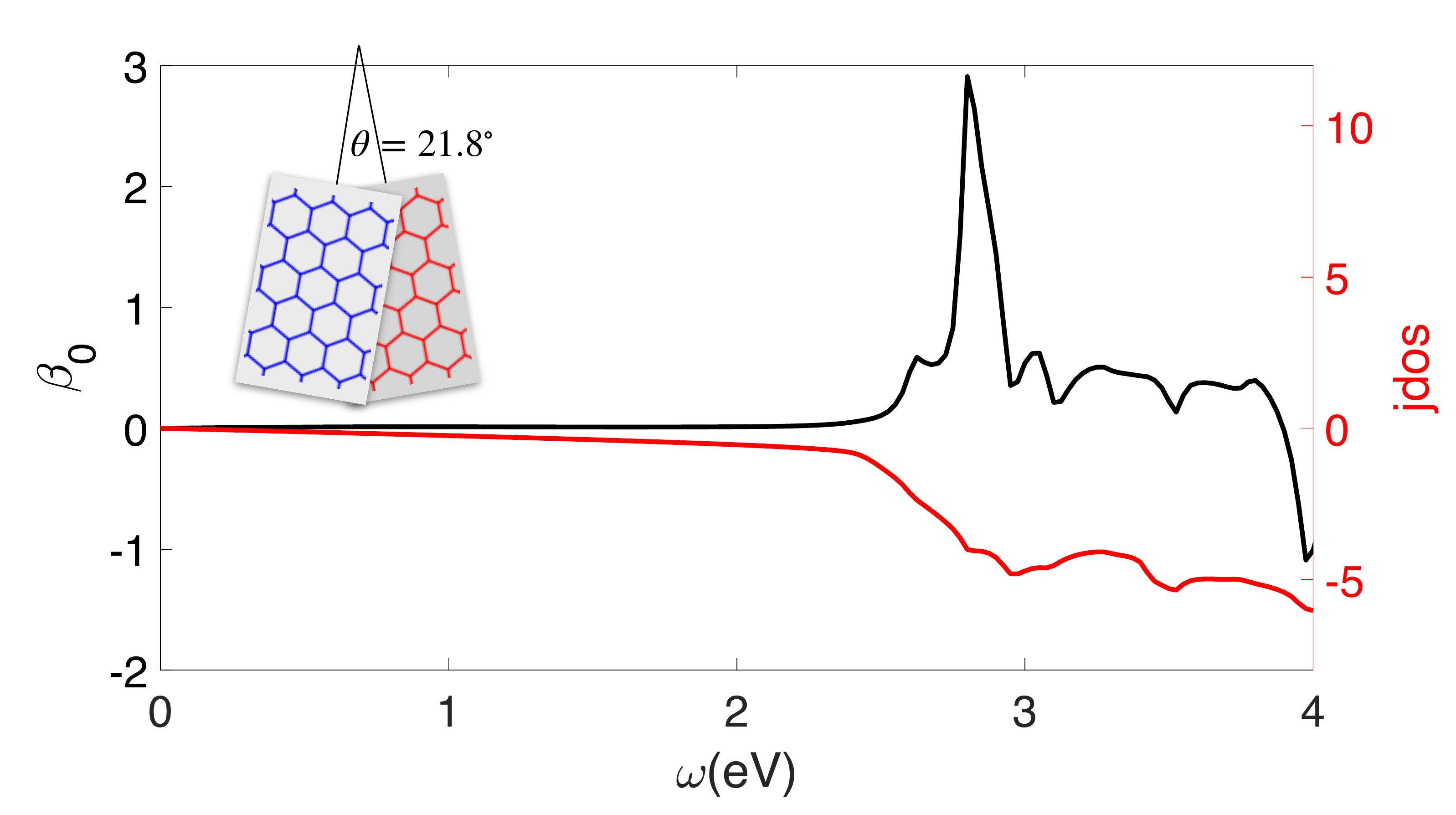}
\caption{The LCPGE coefficient $\beta_0$ and the joint density of states~(jdos) in twisted bilayer graphene with twisting angle $\theta = 21.8^\circ$. $\beta_0$ is in unit of $e/\hbar^2$. The joint density of states is in units of $1/3a^2$ per electron volts with $a$ being the lattice constant of the bilayer graphene. The phenomenological relaxation parameters are chosen as $\eta_1=\eta_2=0.02$ eV. The Fermi level is taken at the Dirac point.}\label{fig_lcpge}
\end{figure}

{\it Twisted bilayer graphene.}---We now apply our theory of the LCPGE to twisted bilayer graphene.  We begin with an AB stacked bilayer graphene, then twist one of the layers around a point where the top and bottom lattice points overlap.  At arbitrary twisting angle except when $\theta = n\pi/3$, the resulting structure, whether commensurate or incommensurate, always respects the chiral $D_3$ or $D_6$ group~\cite{Kang2018,Angeli2018,Zou2018,Zhang2019}.

The twisted bilayer graphene is modeled using a tight-binding Hamiltonian at commensurate angles, following the procedure outlined in Ref.~\cite{Moon2013}.
To see the characteristic behavior of the LCPGE, we choose the twisting angle $\theta=21.8^\circ$, which contains $28$ atoms in the unit cell.  We plot $\beta_0$ as a function of the photon energy in Fig.~\ref{fig_lcpge}, calculated using Eq.~\eqref{eq_res1}.  We can see that below a threshold photon energy, $\beta_0$ is approximately zero. It then develops a sharp peak at around 2.8 eV, followed by finite but oscillating behavior.

\begin{figure}[t]
 \includegraphics[width=\columnwidth]{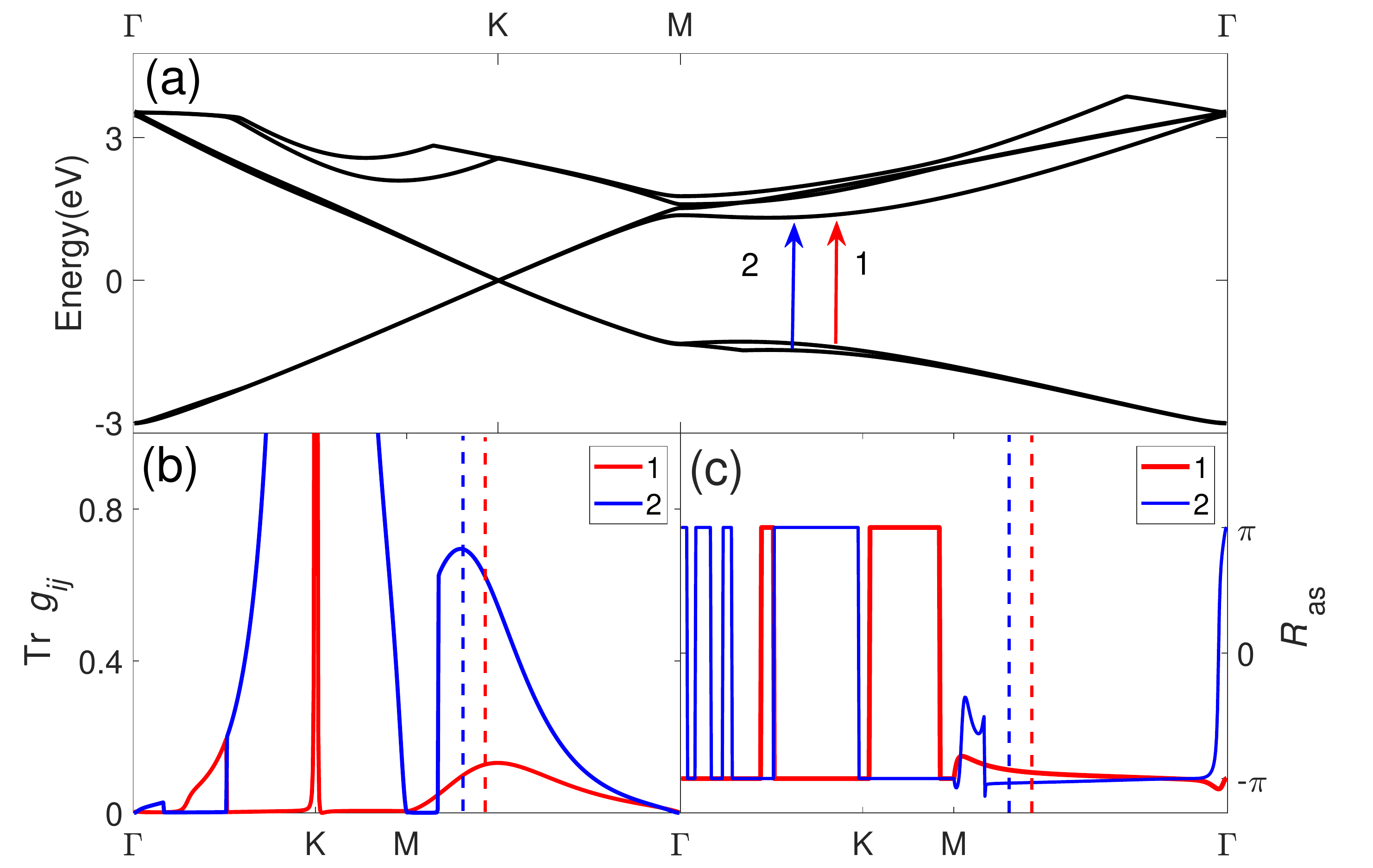}
\caption{Geometric origin of the peak at 2.8 eV in Fig.~\ref{fig_lcpge}. We plot the band spectrum in (a), the quantum geometric tensor $(\TR g_{ij})_{n\ell}$ in (b) and the interlayer coordinate shift $R_\text{as}$ in (c).  The red and blue arrows in (a) show the optical transition with a photon energy 2.8 eV. For $(\TR g_{ij})_{n\ell}$ and $(R_\text{as})_{n\ell}$, we consider the quantities with $n=14$ and $\ell=15$, and with $n=13$ and $\ell=15$, which correspond to the red and blue transitions shown in (a). The blue and red dashed lines in (b) and (c) are at the same position in the Brillouin zone with the blue and red arrows in (a).}\label{fig_geo}
\end{figure}

Three factors conspire to render the appearance of the resonance peak in $\beta_0$: a large joint density of states, a peak in the geometric tensor~($\TR g$), and a finite shift between layers~($R_\text{as}$).
To demonstrate this, we first plot the joint density of states~(jdos) in conjunction with $\beta_0$ in Fig.~\ref{fig_lcpge}, which is defined by
\begin{align}\label{eq_jdos}
{\rm jdos}={\rm Im}\int \frac{d\bm k}{(2\pi)^2} \sum_{m\neq n} \frac{f_m-f_n}{\omega_{mn}-\omega-i\eta_2}\,.
\end{align}
In the flat region where $\beta_0$ is close to zero, the joint density of states varies linearly, demonstrating the typical behaviour of the two-dimensional Dirac point. As the joint density of states rises sharply, so is $\beta_0$. However, the first peak of $\beta_0$ does not coincide with that of the joint density of states, although there are synchronized but much weaker peaks at higher energies.  This shows that even though the increased density of states contributes to the enhanced $\beta_0$, it is not the only factor.

We now reveal the geometric origin of the peak. In Fig.~\ref{fig_geo}(a), we plot the  energy bands along high symmetry lines and label the optical transition responsible for the peak in $\beta_0$.  We can see that the band dispersion around the Dirac point ceases to be linear at the $M$-$\Gamma$ line that bisects the two Dirac points.  On this line a local band edge is developed, rendering a relatively flat region, which hosts the optical transition for the peak in $\beta_0$ and is responsible for the sharp rise of the joint density of states shown in Fig.~\ref{fig_lcpge}.  This will provide ample initial and final states for the optical transitions and hence amplify the magnitude of $\beta_0$.

In Fig.~\ref{fig_geo}(b) and (c), we plot the quantum metric tensor and the layer shift $R_\text{as}$ corresponding to the two transitions labeled in Fig.~\ref{fig_geo}(a). We find that near both transitions, the shift $R_\text{as}$ is close to $-\pi$ without much variation, while the quantum metric tensor experiences peaks which eventually leads to the peak in $\beta_0$.   This turns out to be the dominant geometric contribution to the LCPGE peak since we have found that there is no contribution from the second term in Eq.~\eqref{eq_geo} as the Berry curvature vanishes for the two bands involved in the optical transition.  This is due to the 2D Dirac point with a vanishing band gap.  We expect this contribution to appear in twisted transition metal dichalcogenides.

The appearance of the resonance peak in the LCPGE is a general feature of twisted bilayer graphene. As shown in Fig.~\ref{fig_fig5}, the resonance frequency varies from visible to infrared as the twisting angle decreases from 22$^\circ$ to 2$^\circ$.  This is expected because the twisting angle controls the energy where the Dirac cones from the top and bottom layer intersects, at which the LCPGE becomes appreciable.  Figure~\ref{fig_fig5} also shows the induced voltage difference between the two layers, calculated from $V=\bracket{p}/\epsilon_0$ with $\epsilon_0$ the vacuum permittivity.  This voltage difference can be probed by, for example, capacitance measurement~\cite{Young2011}.  We can see that the voltage increases considerably with decreasing twisting angles, owing to the decreasing peak frequencies and the prefactor $1/\omega^2$ in $\beta$.  The LCPGE signal can be further enhanced in multilayer chiral stacked structures.

\begin{figure}[t]
\includegraphics[width=0.85\columnwidth]{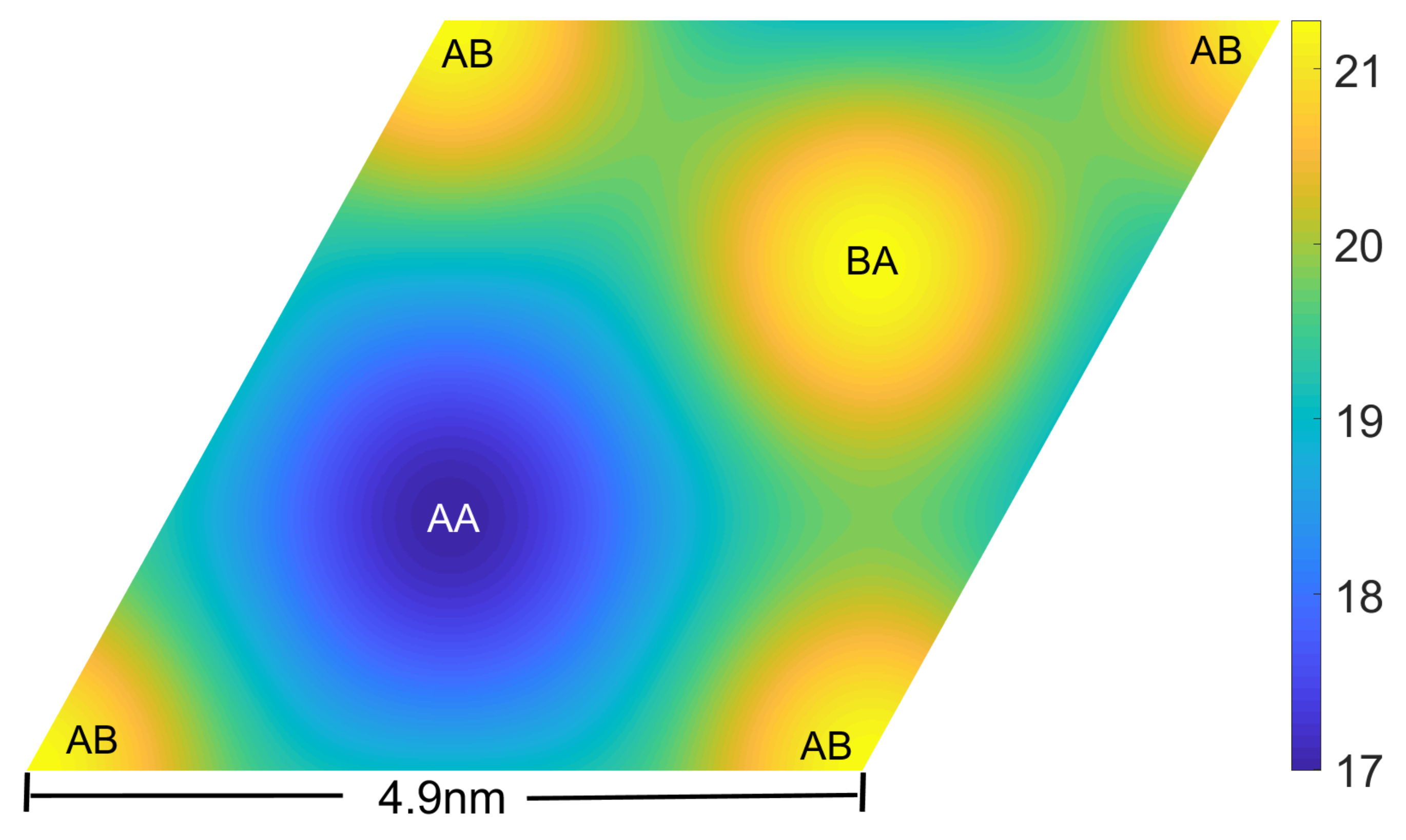}\\
\caption{\label{fig_fig6}The real-space resolved layer potential difference~($\mu$V) induced by the LCPGE in the moire supercell at the twisting angle $\theta=2.87^\circ$ and the peak frequency $\omega=0.34$ eV.}
\end{figure}

Finally, we plot the real-space map of the induced layer potential in Fig.~\ref{fig_fig6} at $\theta = 2.87^\circ$.  We see that the potential is almost uniform throughout the entire unit cell with moderate variation, with minima near the AA-stacked region and maxima near the AB/BA-stacked region.  Note that the potential never changes sign.  Therefore if there is moderate lattice relaxation or strain, we do not expect any cancellation effect, and the LCPGE should be relatively robust effect against relaxation.  On the other hand, it has been shown that for twisting angle smaller than $2^\circ$, the lattice relaxation changes the band structure significantly~\cite{Nam2017}, and we leave the LCPGE in this situation for future study.

In summary, we have studied the layer circular photogalvanic effect in quasi-two-dimensional chiral materials, and revealed its geometric origin.  This geometric view offers a route to designing nonlinear optical chiral materials.  The calculated LCPGE coefficient in twisted bilayer graphene exhibits a highly tunable resonance peak as a function of the twisting angle and photo energy, which may be useful for frequency-sensitive, circularly-polarized light detection, particularly in the infrared range

\begin{acknowledgments}
We acknowledge useful discussions with Dmitri Basov, Kenneth Burch, Ben Hunt, Abhay Pasupathy, and Dong Sun.  We also thank Chong Wang for his suggestions for code optimization. This work is mainly supported by the Department of Energy, Basic Energy Sciences, Grant No.~DE-SC0012509.  The modeling of the moire superlattice is supported by the Department of Energy, Basic Energy Sciences, Pro-QM EFRC (DE-SC0019443).  D.X.\ also acknowledges the support of a Simons Foundation Fellowship in Theoretical Physics.
\end{acknowledgments}

\end{document}